\documentclass{aastex702}
\usepackage{amssymb}
\usepackage{booktabs}
\newcommand{\eps}{e$^-$\,s$^{-1}$}
\newcommand{\epp}{e$^-$\,s$^{-1}$\,pixel$^{-1}$}

\newcommand{\dpd}{$^\circ$\,day$^{-1}$}
\newcommand{\afr}{$Af\rho$}

\shorttitle{Comets in Kepler background photometry}
\shortauthors{Shearer}

\begin{document}

\title{Comets in \emph{Kepler} background photometry: the ion tail and tail disconnection of C/2009~K5
(McNaught) and an analysis of comets crossing the \emph{Kepler} field}

\author{Peter M. Shearer}
\affiliation{Scripps Institution of Oceanography, U.C. San Diego, La Jolla, CA 92037, USA}
\email[show]{pshearer@ucsd.edu}

\begin{abstract}
\emph{Kepler} recorded several thousand dedicated sky-background pixels per detector output every 29.4~min from
2009 to 2013, an archive that we use to study comets. During 2010 April 14--21 a narrow band of
excess light swept across the field at $1.46^\circ$\,day$^{-1}$. This feature is the ion tail of comet C/2009~K5 (McNaught),
whose head passed just beyond the edge of the detector array. We trace the tail from $0.9^\circ$ to $\sim13^\circ$
from the nucleus at surface brightnesses of 26.7--30~mag\,arcsec$^{-2}$; its aberration implies a solar-wind speed
of $\approx500$--750~km\,s$^{-1}$. At 30-min cadence we follow a tail-disconnection event on 2010 April 17--19 whose
timing and geometry are consistent with a flank encounter with a coronal mass ejection, although other triggers
cannot be excluded. We then analyze all 21 known comets with perihelion distance $q<15$~AU whose heads crossed the
detectors, using photometry at the predicted head positions calibrated against displaced control tracks. Eight
comets are detected, including C/2010~J4 (WISE) a few days before its discovery and two comets at 6.3--7.4~AU with
one-sided streaks that we interpret as dust tails. The undetected comets give $3\sigma$ limits of typically
$K_p>16.6$--18.1 within $5'$ of the head. A blind shift-and-stack search for moving diffuse sources, roughly 50\%
complete at $K_p\approx15.5$--16 for rates of 0.3--1\dpd, recovers four known comets and finds no unknown ones.
Animations show the K5 tail and comets C/2011~L3 and C/2010~J4 crossing the field.
\end{abstract}


\section{Introduction}

The \emph{Kepler} mission \citep{Borucki2010,Koch2010} observed a $\sim115$~deg$^2$ field in Cygnus and Lyra almost
continuously from 2009 May to 2013 May, recording the brightness of $\sim1.6\times10^5$ stars every 29.4~min. To
estimate the sky background beneath its targets, \emph{Kepler} also downlinked several thousand dedicated
``background'' pixels per charge-coupled device (CCD) output at every long cadence \citep{Jenkins2010}. The
processing pipeline isolates the brightness variations of the target stars by fitting a smooth two-dimensional surface to these pixels at each cadence and subtracting it from the
target star apertures \citep{Twicken2010}; the background pixels themselves are archived but rarely used directly for science.

While searching the \emph{Kepler} light curves for brightness changes that are correlated among neighboring stars
--- a test for unrecognized instrumental effects as well as for time-varying diffuse foreground structure --- we found a
narrow, moving band of excess light in observing Quarter~5 (Q5). This paper documents that event, identifies it as
the ion tail of comet C/2009~K5 (McNaught), and uses the background pixels to measure the tail's structure and
evolution, including a disconnection event that we associate with a coronal mass ejection (CME). Comet plasma tails
are shaped by the solar wind, and disconnection events --- in which the tail separates from the coma and a new tail
forms --- have been linked to heliospheric current-sheet crossings and to CMEs
\citep[e.g.,][]{Niedner1978,Brandt1992,Vourlidas2007}. Ground-based detections of faint, distant parts of plasma
tails are limited mainly by systematic sky-subtraction and flat-field errors rather than by photon noise.
\emph{Kepler}'s fixed pointing and repeated imaging of the same locations allow each pixel to be compared with its
own history, which removes static sky structure and gives sensitivity to diffuse features as faint as $\sim10^{-3}$ of
the sky background recorded by \emph{Kepler}.

Because the background pixels sample continuously for four years, they contain information about every comet that crossed the \emph{Kepler} field. We systematically consider all known comets whose heads fell
on the \emph{Kepler} detectors, measure the brightness of detected comets, and constrain the maximum brightness of undetected comets.  
We also perform a blind search of \emph{Kepler} data for unknown comets but find none. 
In addition to comets, the background \emph{Kepler} pixels record Thomson-scattered solar-wind structure (P. Shearer 2026, submitted to ApJ), which provides context for the solar-wind interaction seen in the K5 tail.

\section{Data processing}

\subsection{Time, units and magnitudes}
Times are given as Barycentric \emph{Kepler} Julian Date, BKJD $=$ BJD $-\,2454833.0$, where BJD is the Julian Date
in Barycentric Dynamical Time (TDB) corrected to the solar-system barycenter; BKJD~0 is 2009 January 1, 12:00~TDB.
Calendar dates quoted in Universal Time (UT) are approximate to a few minutes. \emph{Kepler} data are naturally grouped into quarters (Q0--Q17), as the spacecraft rolled by
$90^\circ$ four times per year to keep its solar arrays facing the Sun.

Fluxes are in the calibrated units of
the \emph{Kepler} data products, photoelectrons per second (\eps), and surface brightness in \epp\ for the
$3.98''$ pixels. With the \emph{Kepler} zero point ($K_p=12$ corresponds to $1.74\times10^5$~\eps, where $K_p$ is
the \emph{Kepler}-band magnitude), 1~\epp\ corresponds to 28.1~mag\,arcsec$^{-2}$ in the \emph{Kepler} band
($\approx420$--900~nm). The background sky level is 140--280~\epp, so 1~\epp\ is 0.4\%--0.7\% of the sky. Zodiacal light
contributes most of it and all of its seasonal variation, as expected for a field at ecliptic latitude $\approx65^\circ$ whose solar elongation
changes through the year \citep{Leinert1998}. The \emph{Kepler} field is close to the Milky Way, and unresolved
Galactic starlight adds a constant part that
increases toward the Galactic plane \citep{Leinert1998}. From the background levels of the individual CCD outputs, we
find that it rises by $\approx6$~\epp\ per degree of decreasing Galactic latitude $b$ across the field, which spans
$b=6^\circ$--$21^\circ$; this gradient is the same in every quarter, independent of the zodiacal level.

\subsection{Light curves and initial detection}
We used the long-cadence light curves of \emph{Kepler} Data Release 25 (DR25) for all quarters (Q0 to Q17), reorganized into
time-ordered daily files containing every observed star. We used the fluxes corrected
by the Presearch Data Conditioning module (PDCSAP; \citealt{Smith2012,Stumpe2012}), removed a 3-day running mean
from each star, and combined stars within $10'$--$15'$ cells to produce quarterly movies of average star brightness variations
within 3--6\,h time bins. Reviewing the Q5 (BKJD 443.5--538.2) results, we noted a narrow
band of star brightening  moving across the north-eastern CCD modules between BKJD 467 and 472. Some faint stars
($K_p\approx15$--16) on the band apparently brighten by up to $\sim0.4\%$ (5--9$\sigma$ in 1-h bins), but the
typical star on the band appears to brighten by only $\sim10^{-4}$, 1--3~\eps, so the band is relatively faint in maps made from the light curves. The same event is far clearer in the background pixels (Section~2.4), and Figure~\ref{fig:overlay} shows it as seen
there.

\subsection{Target pixel files}
To determine whether the stars or the sky changed, we examined Q5 target pixel files (TPFs), which contain the
individual pixel values around each target star, for 16 stars on the band (CCD channel~71), 8 control stars nearby,
and 6 stars on the band at later times. For each star we detrended every pixel and decomposed the excess image at
the time of brightness maximum into a component proportional to the star's mean image and a spatially uniform pedestal. For all
16 band stars the star-shaped component is consistent with zero ($|z|<1.6$), whereas the uniform pedestal is
1--4~\epp\ (3--10$\sigma$) and appears equally in pixels outside the photometric aperture. Stacking the band stars gives a pedestal lasting $\approx1.5$~h (full width at half
maximum, FWHM) with a peak of $\approx2.8$~\epp. The pipeline background model rose by only $\approx0.5$~\epp\ at
these times: the band is too narrow to be captured by the smooth background surface, so the excess light remains in
the apertures and appears in the light curves. Its fractional effect is largest for faint stars, explaining the
initial detection.

\subsection{Background pixels}
\emph{Kepler} recorded 4200 background pixels per CCD output, arranged as $\sim1050$ groups of $2\times2$ pixels
on a grid with $\approx2'$ spacing. We downloaded all 1460 quarterly background files (84 per quarter in Q0--Q4, 80 per quarter
after the failure of CCD module~3 in 2010 January) from the Mikulski Archive for Space Telescopes (MAST) and reduced
each output to per-cadence surface-fit coefficients and residuals for each $2\times2$ group (12~GB in total). Group
positions were converted to sky coordinates with a cubic map per output that takes the flux-weighted centroids of
$\sim1500$--3500 stars on that output to their catalogued positions (rms residual $<1.1''$). The per-group noise is $\approx1.0$~\epp\ per
cadence.

All comet measurements use the same processing steps. From each group we remove (i) a smooth spatial surface per
output, consisting of a robust plane fitted at every cadence plus cubic terms smoothed over one day, and (ii) a
temporal baseline, a running median over $\pm1$~day for the K5 maps or matched to each comet's angular speed for the
head photometry (Section~\ref{sec:census}). Refitting a full cubic surface at every cadence would absorb much of a tail
$10'$--$20'$ wide, and even the per-cadence plane absorbs some light from a bright structure crossing an output,
producing shallow negative side lobes. For the quantitative measurements of the K5 tail (and for display of the
bright coma of C/2011~L3) we therefore exclude groups within $15'$ of the tail axis or head from the plane fit;
without this, the peak brightness of the tail is underestimated by up to a factor of 1.6 and its width by up to
40\%. For head photometry this is unnecessary, because the side lobes are broad and cancel in the difference between
the head aperture and the surrounding ring.

\subsection{Instrumental effects}
Because all our measurements are relative changes on time scales from 30~min to $\sim10$ days, slow changes of the
zodiacal light are removed by the temporal baselines, and the main concern is instrumental effects on shorter time
scales \citep{VanCleve2016,VanCleve2016b}. We mask cadences flagged by the standard \emph{Kepler} processing pipeline for attitude adjustments, coarse
pointing and reaction-wheel momentum desaturations (every few days), and drop the 5\% of pixel groups with the
largest cadence-to-cadence scatter on each output, which are mostly near bright or variable stars. Thermal
transients after downlinks and rolls, energetic-particle events and electronic glitches change the background over
whole outputs and are largely removed by the per-cadence plane; none of these effects produces a moving, narrow
structure like the K5 tail.

\subsection{Ephemerides}
Positions of comets as seen from \emph{Kepler} were computed with the Jet Propulsion Laboratory (JPL) Horizons system
\citep{Giorgini1996} using observer code 500@$-227$. Horizons also provides the position angle of the extended
Sun--comet radius vector (PsAng, the anti-solar direction, expected for an ion tail) and of the negative of the
comet's heliocentric velocity vector (PsAMV, the direction in which old dust trails); position angles are measured
from north through east. For 2010 April 14--18, C/2009~K5 was at heliocentric distance $r=1.44$--1.43~astronomical
units (AU), distance from \emph{Kepler} $\Delta=1.31$~AU, solar elongation $75^\circ$ and phase angle
$\alpha\approx43^\circ$, moving north at $1.45$\dpd\ just east of the \emph{Kepler} field.

\section{The ion tail of C/2009 K5}

\subsection{Identification}
\begin{figure}[p]
\centering
\begin{interactive}{animation}{c_fig1_anim.mp4}
\includegraphics{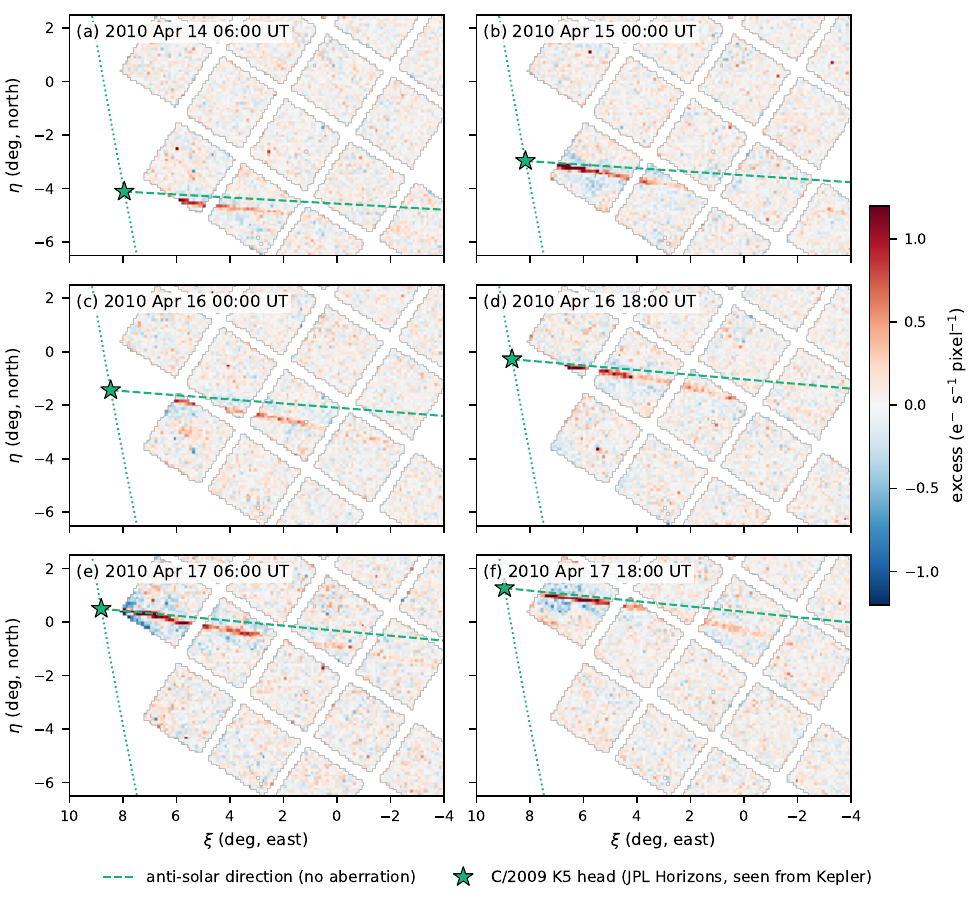}
\end{interactive}
\caption{The ion tail of C/2009~K5 in \emph{Kepler} background pixels at six times (2-h averages, all 80 CCD
outputs, $6'$ cells; tangent plane centered on the \emph{Kepler} field, east left, north up; red brighter). Only the
part of the field crossed by the tail is shown. The star marks the head position predicted by JPL Horizons as seen
from \emph{Kepler} (off the detectors), the dotted line its track, and the dashed line the anti-solar direction.
An animation of this figure is available. It shows the tail from BKJD 465.6 to 475.4 (2010 April 12 02:24 to April 21
21:00 UT) at 1-h steps and runs for 20~s. Left: excess surface brightness on the sky as in the static figure, with
1.5-h averages. Right: the same data in tail-aligned coordinates (distance $L$ from the head; cross-tail offset $s$
from an axis at PsAng$-4.4^\circ$), 3-h averages. The disconnection event begins near BKJD 471.5 (April 18 00~UT).}
\label{fig:overlay}
\end{figure}

In the background pixels the band is a straight line visible in every 1-h frame from BKJD 467.2 to 471.8, and
faintly to 474.8 in the northern modules. In each frame we evaluated the mean excess of background groups in $2'$
strips along straight lines over a grid of orientations, a discrete Hough transform \citep{Hough1962,Duda1972}, and
took the orientation and offset with the highest signal-to-noise ratio. The orientation is constant to $\pm2^\circ$
and the band moves perpendicular to itself at $1.46\pm0.02$\dpd. The predicted head of C/2009~K5 lies on the extension of
the band, $0.7^\circ$--$1^\circ$ beyond the edge of the detector array, and moves with it at the same rate
(Figure~\ref{fig:overlay}). The band lies within $\approx5^\circ$ of the comet's anti-solar direction, offset in the
sense expected for the aberration of a plasma tail (Section~\ref{sec:aberr}); extended back past the head, its great
circle passes $\approx4^\circ$ from the Sun, as expected for a tail $4.4^\circ$ from the anti-solar direction at a
solar elongation of $75^\circ$. The dust-tail direction (PsAMV $\approx201^\circ$) points away from the field. We
conclude that the band is the plasma tail of C/2009~K5. The animated version of Figure~\ref{fig:overlay} shows the tail from BKJD 465.6 to 475.4.

\subsection{Tail structure}
\label{sec:structure}
We transformed every background group and cadence into tail-aligned coordinates: projected distance $L$ from the
nucleus and cross-tail offset $s$ from an axis at PsAng$-4.4^\circ$. The tail is detected from $L=0.9^\circ$ (the
nearest part of the detectors) to $\approx13^\circ$. Its axis lies $4.4^\circ\pm0.3^\circ$ from the projected
anti-solar direction for $L<6^\circ$, increasing to $\approx5.0^\circ$--$5.3^\circ$ at $L=7^\circ$--$8.6^\circ$, so
the tail curves slightly backward. The FWHM increases from $6'$ at $L\approx1^\circ$ to $12'$--$14'$ at
$1.5^\circ$--$2.5^\circ$ and $\approx20'$ beyond $6^\circ$ (projected widths of $3\times10^5$ to
$\approx1.2\times10^6$~km at $\Delta=1.31$~AU). The tail is brightest within $\approx3^\circ$ of the head, where the
peak surface brightness is 2--3.6~\epp\ (26.7--27.4~mag\,arcsec$^{-2}$; 1.5\% of the sky at $L=0.9^\circ$). It falls
to $\approx0.6$~\epp\ by $L\approx3.5^\circ$ and to 0.2--0.4~\epp\ ($\approx29$--30~mag\,arcsec$^{-2}$) beyond
$5^\circ$--$6^\circ$. These peak values change by less than a factor of 1.5 between analysis choices; the brightness
integrated across the tail is more sensitive to how the widening outer tail is separated from the background, and
we do not attempt to fit a functional form to it.

The brightness varies with time by tens of percent, largely simultaneously along the visible tail and partly as
the tail swings across its mean axis; we find no discrete condensations moving down the tail before the
disconnection. In the \emph{Kepler} band the tail light is presumably a mixture of emission from CO$^+$, whose comet-tail bands are
strongest at 400--500~nm where \emph{Kepler}'s response is low, and H$_2$O$^+$, whose bands at 550--750~nm
\citep{Herzberg1974} fall where the response is high. The broad band does not allow the two to be separated.

\subsection{The aberration angle}
\label{sec:aberr}
A plasma tail lies along the solar-wind velocity in the comet's frame, $w\hat{\mathbf r}-\mathbf v_c$, where $w$ is
the (assumed radial) solar-wind speed and $\mathbf v_c$ the comet's heliocentric velocity, and so is aberrated from
the anti-solar direction by an angle that depends on $w$ \citep[e.g.,][]{Brandt1967,Buffington2008,Li2023}. We computed the comet's
heliocentric position and velocity from the Horizons ephemeris (35~km\,s$^{-1}$, almost entirely perpendicular to the
radius vector) and projected $w\hat{\mathbf r}-\mathbf v_c$ onto the sky as seen from \emph{Kepler}. The projected
direction for $w\to\infty$ reproduces the Horizons PsAng to $0.1^\circ$. The predicted aberration is $-8.6^\circ$,
$-6.5^\circ$, $-5.3^\circ$, $-4.4^\circ$, $-3.6^\circ$ and $-2.7^\circ$ for $w=300$, 400, 500, 600, 750 and
1000~km\,s$^{-1}$, with the same sign as observed. The measured $4.4^\circ\pm0.3^\circ$ near the head therefore
implies $w\approx600$~km\,s$^{-1}$ (statistical uncertainty $\pm50$~km\,s$^{-1}$), and $\approx5^\circ$ farther out
implies $\approx530$~km\,s$^{-1}$. The systematic uncertainty is larger. The method assumes a radial wind and a tail
lying along the local wind direction, whereas non-radial flow of a few degrees, the backward curvature of the tail
and the strong projection (the tail lies $\approx137^\circ$ from the line of sight) all affect the measured angle.
An error of $1^\circ$ in the aberration corresponds to $w\approx500$--750~km\,s$^{-1}$, and we adopt this range. The
comet was $\approx51^\circ$ north of the solar equator, where fast wind from the polar coronal hole
($\sim750$~km\,s$^{-1}$) would be expected near solar minimum \citep{McComas2000,McComas2008}; the inferred speed is intermediate between typical slow
and fast wind.

\subsection{A tail-disconnection event}
\begin{figure}[p]
\centering
\includegraphics{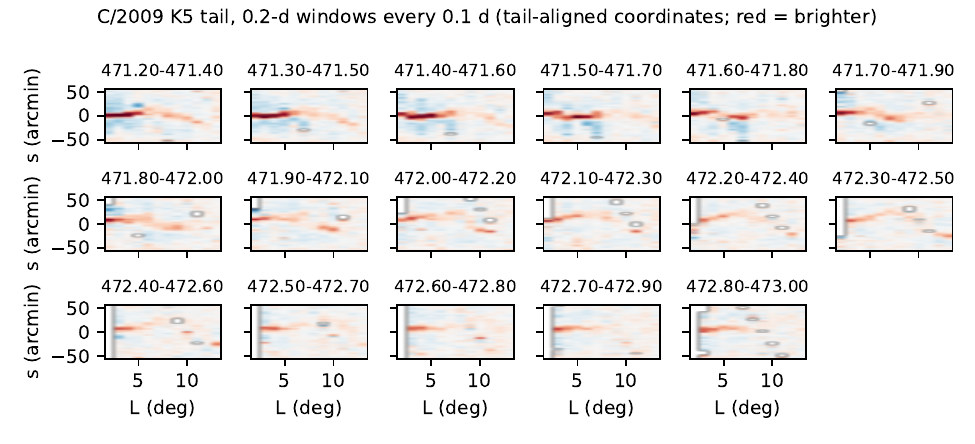}
\caption{The disconnection event: images in tail-aligned coordinates (horizontal: distance $L$ from the head;
vertical: cross-tail offset $s$ from an axis at PsAng$-4.4^\circ$) in 0.2-day windows every 0.1~day, BKJD
471.2--473.0, all 80 outputs.}
\label{fig:kink}
\end{figure}

At 0.2-day resolution using all 80 outputs (Figure~\ref{fig:kink}), the tail is straight and bright until BKJD
471.5. Between 471.5 and 471.7 a gap opens at $L\approx4^\circ$--$5^\circ$ and the tail beyond it is displaced by
$\approx10'$. The detached segment then recedes from the nucleus and drifts sideways: it lies at
$L\approx7.5^\circ$ near BKJD 471.8, $\approx10^\circ$ near 472.0 and $12^\circ$--$13^\circ$ by 472.4--472.6, where
it is displaced $\approx25'$ from the axis. Meanwhile the inner tail swings $15'$--$20'$ to the other side, and by
BKJD 472.5--473.0 a new straight tail has formed, offset $\approx5'$ from the earlier axis and brightest near the
nucleus. The detached segment moves at $\approx7$\dpd, or $\approx280$~km\,s$^{-1}$ projected on the sky
($1^\circ=3.4\times10^6$~km at $\Delta$). If the tail lies at $\approx137^\circ$ from the line of sight, as expected
for its aberrated direction, the speed along the tail is $\approx400$~km\,s$^{-1}$; for angles of $130^\circ$--$145^\circ$
it would be $\approx370$--490~km\,s$^{-1}$. The sequence (break near the head, recession of the old tail,
regrowth) is characteristic of a disconnection event \citep[e.g.,][]{Niedner1978,Vourlidas2007,Kuchar2008}.

The event overlaps a \emph{Kepler} momentum desaturation (BKJD 471.79) and reaction-wheel zero crossings (471.98--472.18;
Section~2.5). An instrumental origin is unlikely: the break develops by 471.5--471.7, before these operations; the
deformation depends on distance from the nucleus and is coherent across many CCD outputs as the tail crosses them,
whereas pointing jitter would act uniformly; and diffuse sky light is insensitive to small pointing changes.

\subsection{Association with a coronal mass ejection}
On 2010 April 18 the comet was at heliocentric ecliptic longitude $238^\circ$ and latitude $+54^\circ$, about
$30^\circ$ west of the Sun--Earth line and $\approx51^\circ$ north of the solar equator in Heliocentric Earth
Equatorial (HEEQ) coordinates. No in-situ solar-wind measurements exist at that location: the near-Earth monitors
and the two Solar Terrestrial Relations Observatory (STEREO) spacecraft were near the ecliptic,
$38^\circ$--$100^\circ$ away in longitude, and \emph{Ulysses} had ceased operation.

The CME catalog of the Large Angle and Spectrometric Coronagraph (LASCO) on the Solar and Heliospheric Observatory
(SOHO; \citealt{Yashiro2004,Gopalswamy2009}) lists one substantial CME in the plausible launch window (April 11--15)
directed toward the comet's position angle as seen from Earth ($\approx335^\circ$--$345^\circ$): first seen on 2010
April 13 at 08:30~UT, central position angle $321^\circ$, apparent width $120^\circ$, linear speed
452~km\,s$^{-1}$ (second-order speed at $20\,R_\odot$ of 709~km\,s$^{-1}$), mass $4.9\times10^{15}$~g. The same CME
appears in the HIGeoCAT catalog of CMEs observed by the STEREO heliospheric imagers, compiled by the Heliospheric
Cataloguing, Analysis and Techniques Service (HELCATS; \citealt{Harrison2018,Barnes2019}):
\begin{center}
\begin{tabular}{lcccc}
\toprule
Spacecraft & Speed (km\,s$^{-1}$) & HEEQ lon & HEEQ lat & Launch (UT, 2010 Apr 13)\\
\midrule
STEREO-A & 502--516 & $+26^\circ$ & $+26^\circ$ & 06:10--07:49\\
STEREO-B & 527--530 & $-5^\circ$ & $+21^\circ$ & 09:00--09:10\\
\bottomrule
\end{tabular}
\end{center}
The ranges span the three geometric fitting methods used in the catalog (fixed-$\phi$, self-similar expansion and
harmonic mean). At constant speed, a CME launched on April $13.3$ at 510--530~km\,s$^{-1}$ reaches 1.43~AU on April
$18.1\pm0.2$ (BKJD $471.6\pm0.2$). The flank of a CME arrives later than its apex, and CME speeds change in transit,
so the arrival at the comet is uncertain by about a day (roughly April 17.6--19). The tail disturbance began by April 18.0--18.2 (BKJD 471.5--471.7) at $L=4^\circ$--$5^\circ$,
implying arrival at the nucleus on April $\approx17.5$--18.1. The comet lay $25^\circ$ (STEREO-A fit) to $40^\circ$
(STEREO-B fit) from the fitted CME apex, comparable to the CME half-width, so the encounter would have involved the
CME flank \citep[cf.][for a flank encounter at Earth on 2010 April 5]{Mostl2010}. No other catalogued CME in the
window is directed toward the comet. We therefore propose that the April 13 CME triggered the disconnection, which has similarities to those
observed for comet 2P/Encke by \citet{Vourlidas2007} and for C/2001~Q4 (NEAT) with the Solar Mass Ejection Imager by
\citet{Kuchar2008}. The association rests on timing and direction alone. Disconnection events have also been
attributed to crossings of the heliospheric current sheet \citep{Niedner1978} and to compressions at stream
interfaces, and without solar-wind measurements near the comet we cannot exclude these. Images of the CME front
near the comet from the STEREO heliospheric imagers would provide a direct test.

\subsection{Tails of comets whose heads did not cross the field}
For all known comets in our sample (Section~\ref{sec:census}) we identified times when the anti-solar or dust-tail
direction from a comet outside the field passed within $8^\circ$ of the field center. A blind search of all quarters
for straight, moving bands in the star light curves recovers the K5 tail and no other band. A targeted stack along
the predicted tail direction gives no detection for C/2009~P1 (Garradd; tail crossing $34^\circ$ from the nucleus,
$<28\%$ of the K5 tail signal, $3\sigma$), C/2009~O2 (Catalina; $27^\circ$, $<37\%$) and C/2011~UF305 (LINEAR;
$8.5^\circ$, $<32\%$). C/2010~F4 (Machholz) could not be tested because of instrumental disturbances at the end of
Q4.

\section{Known comets crossing the \emph{Kepler} field}
\label{sec:census}

\subsection{Selection criteria}
We selected comets by their orbits rather than by predicted brightness. JPL's magnitude parameters are missing for
more than half of all comets, and for those we detect they underestimate the light within a few arcminutes by up to
3~mag (Section~\ref{sec:obs}), so a brightness cutoff could have removed comets that are in fact detectable. We
retrieved from the JPL Small-Body Database all comets with perihelion distance $q<15$~AU: every periodic comet
($P<200$~yr), and every long-period comet whose perihelion passage fell within 1500~days of the mission
($q<5$~AU) or within 4000~days ($q\ge5$~AU, since distant comets stay near perihelion distance for years). For
each we computed daily Horizons ephemerides as seen from \emph{Kepler}. Twenty-one comets had their heads on the
detectors during \emph{Kepler} observations (Table~\ref{tab:census}; Figure~\ref{fig:paths}), at heliocentric
distances from 1.1 to 12.9~AU and for between 2 and 273 days. Because
\emph{Kepler} was far from the Earth and the comets were mostly distant, their paths across the field are strongly
curved by parallax from \emph{Kepler}'s own orbital motion. The field contains gaps between CCDs and modules, so a
comet can be within the field without being recorded.

\subsection{Head photometry}
For each comet and quarter we selected all background groups within $1^\circ$ of its track, removed the smooth
spatial surface (Section~2.4), masked flagged cadences and the 5\% noisiest groups, and subtracted a running median
in time whose half-width is matched to the comet's angular speed $\dot\theta$: $\mathrm{clip}(6R/\dot\theta,\,2,\,10)$
days for a measurement radius $R=8'$. This preserves slowly moving distant comets, which a fixed short baseline would
partly subtract. In 0.5-day bins we took the mean excess within $R$ of the predicted head minus the mean in a
$2R$--$4R$ ring, evaluating the head position at every cadence (a comet moving at $1$\dpd\ travels $30'$ in a bin),
and averaged the bins weighted by the number of groups within $R$; uncertainties come from the scatter between bins.
Only bins in which the predicted head lies on the detectors (from the daily ephemeris) are used: bins before a
comet arrives or after it leaves sample partly covered regions at CCD edges.
For a coma with surface brightness $S(\rho)=S_1/\rho$, where $\rho$ is the angular distance from the head (in arcmin)
and $S_1$ the surface brightness at $\rho=1'$, this excess equals $S_1[2/R-1/(3R)]=5S_1/(3R)$, which we convert to
$S_1$, to the magnitude within $5'$ ($K_{p,5'}$), and to the dust-production proxy \afr\ (albedo times filling factor
times the aperture radius $\rho$ expressed as a distance at the comet; \citealt{AHearn1984}), using $K_p(\odot)=-26.9$.
The $1/\rho$ form is the projection of steady, isotropic outflow, whose density falls as $r^{-2}$ from the nucleus.
Its central singularity is harmless: the light within an aperture of radius $\rho$, $2\pi S_1\rho$, is finite and
grows linearly with $\rho$. Real comae depart from $1/\rho$ within tens of kilometers of the nucleus and steepen at
large distances, but our measurements sample $\rho\gtrsim1'$ ($\gtrsim10^4$~km), and the innermost $0.5'$
supplies only 10\% of the light within $5'$. Because \afr\ is independent of $\rho$ for a
$1/\rho$ coma, the result does not depend on $R$ to the extent that the profile follows $1/\rho$. Light from a tail
within $8'$ of the head is included, so for comets with tails (C/2007~Q3 and C/2006~Q1; Section~\ref{sec:results}) \afr\ is an upper bound on
that of the coma alone.

To calibrate the significance empirically, we repeated every measurement along eight control tracks: the comet's
own track displaced by $2.5^\circ$ in eight directions on the sky, with the same timing, baseline and sampling but no
comet. Of the 334 control tracks with at least five bins, none exceeds $4.4\sigma$ in absolute value and 3.6\% exceed
$3\sigma$. Their robust scatter depends on speed: 1.48 for tracks slower than $0.15$\dpd\ (189 tracks), but 0.89 and
0.74 for tracks at 0.15--0.5\dpd\ and faster. Slow tracks dwell for weeks on the same sky, where slowly varying stars and residual
instrumental structure add correlated noise that the bin-to-bin scatter does not capture. We therefore divide the
significance of slow comets by 1.48 and, conservatively, leave that of faster comets unchanged. A measurement with
calibrated significance $\sigma_{\rm cal}\ge5$ is a detection and one with $3\le\sigma_{\rm cal}<5$ is marginal.
Because 3.6\% of control tracks exceed $3\sigma$, one or two of the 50 measurements in
Table~\ref{tab:census} are expected to be marginal by chance. Selecting only times of low solar-wind activity, as
measured by the companion paper's field-wide brightness fluctuations, reduces the scatter of the slow control tracks
only slightly (from 1.44 to 1.28 for the bins used in this test) while discarding half of the data, and raises the significance of no detection
appreciably, so the solar-wind signal does not limit the head photometry; the head-minus-ring difference removes most
of it.

\subsection{Results}
\label{sec:results}
\begin{table*}[t]
\centering\small
\caption{Analysis of comets whose heads fell on the \emph{Kepler} detectors (2009--2013): uniform head photometry. Excess is the mean background-pixel signal within $8'$ of the predicted head minus a $16'$--$32'$ ring; $S_1$ is the equivalent $1/\rho$ coma brightness at $\rho=1'$; $K_{p,5'}$ the magnitude within $5'$; upper limits are $3\sigma$. $\sigma_{\rm cal}$ is the significance divided by the scatter of control tracks displaced $2.5^\circ$ from each comet's path (1.48 for tracks slower than $0.15^\circ$/d, 1.0 otherwise). Detections have $\sigma_{\rm cal}\ge5$, marginal measurements $3\le\sigma_{\rm cal}<5$.}
\label{tab:census}
\setlength{\tabcolsep}{3pt}\resizebox{\textwidth}{!}{\begin{tabular}{lrrrrrrrrrl}
\toprule
Comet & Q & days & $r$ (AU) & $\Delta$ (AU) & $\dot\theta$ ($^\circ$/d) & $T_{\rm JPL}$ & excess (\epp) & $\sigma_{\rm cal}$ & $K_{p,5'}$ & $Af\rho$ (cm) \\
\midrule
C/2010 J4 \,(WISE) & 5 & 4 & 1.09 & 0.35 & 5.82 & 17.7 & 0.379$\pm$0.022 & 17.5 & 14.8 & 0.35$\pm$0.02 \\
C/2010 G1 \,(Boattini) & 4 & 28 & 1.81 & 1.90 & 0.51 & 16.6 & 0.136$\pm$0.014 & 9.7 & 15.9 & 1.9$\pm$0.2 \\
C/2011 L3 \,(McNaught) & 10 & 16 & 1.93 & 1.24 & 0.93 & 16.5 & 1.119$\pm$0.062 & 18.1 & 13.6 & 12$\pm$1 \\
C/2012 K5 \,(LINEAR) & 14 & 12 & 2.43 & 1.88 & 0.73 & 16.4 & 0.246$\pm$0.021 & 11.5 & 15.3 & 6.1$\pm$0.5 \\
C/2012 K5 \,(LINEAR) & 13 & 8 & 2.56 & 2.01 & 0.67 & 16.7 & 0.165$\pm$0.012 & 13.7 & 15.7 & 4.8$\pm$0.4 \\
C/2009 Y1 \,(Catalina) & 7 & 4 & 2.58 & 2.47 & 0.73 & 14.2 & 1.131$\pm$0.118 & 9.6 & 13.6 & 42$\pm$4 \\
P/2012 NJ \,(La Sagra) & 15 & 4 & 2.89 & 3.08 & 0.26 & 19.3 & -0.002$\pm$0.010 & -0.2 & $>17.6$ & $<1.8$ \\
P/2012 NJ \,(La Sagra) & 16 & 35 & 3.08 & 3.27 & 0.29 & 19.6 & -0.008$\pm$0.007 & -1.0 & $>17.9$ & $<1.5$ \\
C/2012 L1 \,(LINEAR) & 13 & 26 & 3.53 & 3.65 & 0.41 & 19.1 & 0.003$\pm$0.008 & 0.4 & $>17.6$ & $<2.8$ \\
C/2012 S3 \,(PANSTARRS) & 15 & 28 & 3.68 & 3.57 & 0.41 & 19.0 & 0.027$\pm$0.005 & 5.1 & 17.7 & 2.9$\pm$0.6 \\
C/2012 E3 \,(PANSTARRS) & 9 & 17 & 3.85 & 3.92 & 0.26 & 18.7 & -0.004$\pm$0.004 & -1.1 & $>18.5$ & $<1.6$ \\
C/2012 E3 \,(PANSTARRS) & 8 & 67 & 3.91 & 4.27 & 0.12 & 19.0 & -0.002$\pm$0.005 & -0.3 & $>17.9$ & $<3.2$ \\
C/2012 E3 \,(PANSTARRS) & 7 & 79 & 4.14 & 4.13 & 0.19 & 19.1 & -0.001$\pm$0.006 & -0.1 & $>18.1$ & $<3.0$ \\
C/2011 K1 \,(Schwartz-Holvorcem) & 6 & 27 & 4.19 & 3.61 & 0.40 & 20.7 & -0.004$\pm$0.006 & -0.6 & $>18.0$ & $<2.7$ \\
C/2008 J5 \,(Garradd) & 0 & 5 & 4.74 & 4.50 & 0.27 & 25.4 & -0.011$\pm$0.010 & -1.1 & $>17.6$ & $<6.5$ \\
C/2008 J5 \,(Garradd) & 1 & 34 & 4.91 & 4.53 & 0.26 & 25.6 & 0.012$\pm$0.007 & 1.7 & $>17.5$ & $<8.0$ \\
C/2010 G3 \,(WISE) & 5 & 32 & 4.91 & 4.90 & 0.29 & 18.6 & 0.018$\pm$0.006 & 3.2 & 18.1 & 4.7$\pm$1.5 \\
C/2012 L2 \,(LINEAR) & 11 & 92 & 6.14 & 6.00 & 0.12 & 20.1 & 0.005$\pm$0.009 & 0.4 & $>17.2$ & $<22.1$ \\
C/2007 Q3 \,(Siding Spring) & 9 & 74 & 6.26 & 6.01 & 0.10 & 17.1 & 0.094$\pm$0.008 & 7.5 & 16.3 & 49$\pm$4 \\
C/2007 Q3 \,(Siding Spring) & 10 & 10 & 6.59 & 6.13 & 0.11 & 17.2 & 0.098$\pm$0.013 & 5.2 & 16.3 & 58$\pm$7 \\
C/2006 Q1 \,(McNaught) & 5 & 64 & 6.86 & 6.56 & 0.11 & 17.4 & 0.148$\pm$0.012 & 8.6 & 15.8 & 102$\pm$8 \\
C/2008 J5 \,(Garradd) & 4 & 90 & 6.89 & 7.24 & 0.15 & 29.1 & 0.002$\pm$0.008 & 0.3 & $>17.7$ & $<19.9$ \\
C/2011 Q1 \,(PANSTARRS) & 7 & 56 & 6.95 & 6.80 & 0.21 & 20.1 & 0.004$\pm$0.013 & 0.3 & $>17.2$ & $<32.3$ \\
C/2009 S3 \,(Lemmon) & 7 & 67 & 7.03 & 6.80 & 0.12 & 19.7 & 0.005$\pm$0.003 & 1.0 & $>18.1$ & $<14.4$ \\
C/2009 S3 \,(Lemmon) & 6 & 47 & 7.19 & 6.74 & 0.07 & 19.7 & 0.011$\pm$0.006 & 1.4 & $>17.4$ & $<27.9$ \\
C/2008 J5 \,(Garradd) & 5 & 23 & 7.32 & 7.39 & 0.16 & 29.6 & -0.009$\pm$0.005 & -1.7 & $>18.2$ & $<14.3$ \\
C/2006 Q1 \,(McNaught) & 6 & 90 & 7.42 & 6.96 & 0.09 & 17.6 & 0.127$\pm$0.014 & 5.9 & 16.0 & 109$\pm$12 \\
C/2013 G6 \,(Lemmon) & 10 & 47 & 7.43 & 6.90 & 0.17 & 34.8 & -0.003$\pm$0.006 & -0.4 & $>18.0$ & $<16.5$ \\
C/2013 G6 \,(Lemmon) & 9 & 97 & 7.97 & 8.01 & 0.10 & 36.0 & -0.023$\pm$0.012 & -1.3 & $>16.9$ & $<60.6$ \\
C/2012 L2 \,(LINEAR) & 8 & 67 & 8.37 & 8.80 & 0.13 & 21.8 & 0.008$\pm$0.006 & 0.9 & $>17.4$ & $<47.7$ \\
C/2007 H2 \,(Skiff) & 2 & 3 & 8.67 & 8.09 & 0.09 & 28.7 & 0.043$\pm$0.018 & 1.6 & $>16.0$ & $<167.0$ \\
C/2012 L2 \,(LINEAR) & 7 & 89 & 9.07 & 9.01 & 0.09 & 22.1 & -0.001$\pm$0.004 & -0.1 & $>18.1$ & $<29.4$ \\
C/2013 G6 \,(Lemmon) & 7 & 89 & 9.28 & 9.24 & 0.10 & 38.1 & 0.001$\pm$0.004 & 0.2 & $>18.1$ & $<33.4$ \\
C/2012 L2 \,(LINEAR) & 6 & 90 & 9.73 & 9.32 & 0.09 & 22.3 & 0.006$\pm$0.008 & 0.5 & $>17.2$ & $<81.9$ \\
C/2013 G6 \,(Lemmon) & 6 & 45 & 9.75 & 9.23 & 0.12 & 38.6 & -0.027$\pm$0.027 & -0.7 & $>16.1$ & $<236.9$ \\
C/2006 Q1 \,(McNaught) & 10 & 94 & 9.93 & 9.41 & 0.08 & 18.9 & -0.007$\pm$0.010 & -0.5 & $>17.1$ & $<92.1$ \\
C/2012 V2 \,(LINEAR) & 7 & 53 & 9.95 & 9.70 & 0.08 & 25.3 & 0.005$\pm$0.005 & 0.7 & $>17.7$ & $<58.9$ \\
C/2006 Q1 \,(McNaught) & 11 & 41 & 10.38 & 10.11 & 0.08 & 19.1 & 0.032$\pm$0.005 & 4.3 & 17.5 & 78$\pm$12 \\
C/2012 L2 \,(LINEAR) & 4 & 90 & 11.02 & 11.52 & 0.10 & 23.1 & -0.002$\pm$0.009 & -0.2 & $>17.3$ & $<125.0$ \\
C/2014 W3 \,(PANSTARRS) & 5 & 6 & 11.14 & 11.26 & 0.11 & 24.6 & 0.002$\pm$0.008 & 0.2 & $>17.3$ & $<116.6$ \\
C/2014 W3 \,(PANSTARRS) & 4 & 90 & 11.38 & 11.69 & 0.10 & 24.8 & -0.005$\pm$0.005 & -0.7 & $>17.9$ & $<75.1$ \\
C/2012 L2 \,(LINEAR) & 3 & 89 & 11.64 & 11.64 & 0.07 & 23.3 & 0.015$\pm$0.017 & 0.6 & $>16.4$ & $<318.9$ \\
C/2012 V2 \,(LINEAR) & 4 & 90 & 11.73 & 12.17 & 0.09 & 26.7 & 0.014$\pm$0.016 & 0.6 & $>16.4$ & $<318.3$ \\
C/2012 L2 \,(LINEAR) & 2 & 89 & 12.23 & 11.75 & 0.07 & 23.4 & 0.082$\pm$0.014 & 4.1 & 16.5 & 322$\pm$55 \\
C/2012 V2 \,(LINEAR) & 3 & 89 & 12.33 & 12.24 & 0.07 & 26.9 & 0.015$\pm$0.008 & 1.3 & $>17.0$ & $<210.1$ \\
C/2014 W3 \,(PANSTARRS) & 1 & 34 & 12.61 & 12.33 & 0.09 & 25.4 & -0.018$\pm$0.008 & -1.4 & $>17.4$ & $<155.7$ \\
C/2012 L2 \,(LINEAR) & 1 & 34 & 12.66 & 12.43 & 0.07 & 23.7 & 0.015$\pm$0.005 & 1.9 & $>17.3$ & $<165.6$ \\
C/2014 W3 \,(PANSTARRS) & 0 & 10 & 12.72 & 12.55 & 0.09 & 25.4 & -0.008$\pm$0.006 & -0.9 & $>17.6$ & $<130.0$ \\
C/2012 V2 \,(LINEAR) & 2 & 47 & 12.78 & 12.36 & 0.07 & 27.1 & -0.011$\pm$0.009 & -0.9 & $>17.3$ & $<173.8$ \\
C/2012 L2 \,(LINEAR) & 0 & 10 & 12.80 & 12.76 & 0.07 & 23.7 & 0.010$\pm$0.014 & 0.4 & $>16.6$ & $<341.9$ \\
\bottomrule
\end{tabular}}
\end{table*}

Table~\ref{tab:census} lists the measurements. Figure~\ref{fig:census} shows, for every detected and marginal
measurement in order of significance, a co-moving stack in which the background groups are averaged in the comet's
frame over the interval; the corresponding head light curves are shown in Figure~\ref{fig:lc}.  The comet images in Figure~\ref{fig:census} contain negative side-lobe artifacts, which appear in blue around and along the
tracks of the brighter comets. They come from two steps of the processing (Section~2.4). The plane fitted at every
cadence to each CCD output absorbs part of the coma light, and so is subtracted with a slight excess from the groups
around the comet; and the running-median baseline of each group includes times when the comet was nearby, so it is
slightly too high for groups along the comet's track. The plane lobes are broad and largely cancel in the
head-minus-ring photometry; the baseline effect can reduce the measured excess slightly, so our \afr\ values may be
somewhat underestimated for slow, bright comets.

Eight comets are
detected in their comae (in eleven quarter-long intervals), and three further measurements are marginal:
\begin{itemize}
\item C/2010~J4 (WISE) at 1.09~AU, only 0.35~AU from \emph{Kepler}, $\sigma_{\rm cal}=17.5$, \afr\ $=0.35\pm0.02$~cm
($K_{p,5'}=14.8$);
\item at 1.8--2.6~AU, C/2010~G1 (\afr\ $=1.9$~cm), C/2011~L3 (12~cm), C/2012~K5 (5 and 6~cm in two quarters) and
C/2009~Y1 (42~cm), all at $\sigma_{\rm cal}=9.6$--18;
\item C/2012~S3 (PANSTARRS) at 3.7~AU, $\sigma_{\rm cal}=5.1$, \afr\ $=2.9\pm0.6$~cm ($K_{p,5'}=17.7$, the faintest
detection);
\item C/2007~Q3 (Siding Spring), detected at 6.3 and 6.6~AU in Q9 and Q10 ($\sigma_{\rm cal}=7.5$ and 5.2,
\afr\ $=49\pm4$ and $58\pm7$~cm);
\item C/2006~Q1 (McNaught), detected at 6.9 and 7.4~AU in Q5 and Q6 ($\sigma_{\rm cal}=8.6$ and 5.9, \afr\
$=102\pm8$ and $109\pm12$~cm);
\item C/2006~Q1 at 10.4~AU in Q11 ($\sigma_{\rm cal}=4.3$, \afr\ $=78\pm12$~cm), C/2010~G3 (WISE) at 4.9~AU in Q5
($3.2$, $4.7\pm1.5$~cm) and C/2012~L2 (LINEAR) at 12.2~AU in Q2 ($4.1$), all marginal and, as discussed below,
unconfirmed.
\end{itemize}
The stacks make the ranking visible: the comets at $\sigma_{\rm cal}\gtrsim9$ appear as compact, saturated heads,
while those at 4--6$\sigma$ are fainter but still centered on the predicted position. Because C/2012~S3 lies close to
the detection threshold, we repeated its measurement along 60 further control tracks displaced by $0.5^\circ$--$3^\circ$
in 12 directions; all lie within $\pm1.5\sigma$. For the two distant comets the measurements in different quarters
are independent (different spacecraft roll angles and CCD outputs) and agree in \afr; combined, C/2007~Q3 (Q9 and
Q10) and C/2006~Q1 (Q5 and Q6) are detected at $\approx9\sigma$ and $\approx10\sigma$. All four comets found by the
blind search (Section~\ref{sec:blind}) are among the detections; the K5 ion tail is the ninth comet seen.
C/2006~Q1 is not detected in Q10 (excess $-0.007\pm0.010$~\epp\ at 9.9~AU), whereas interpolating between its Q6 and Q11
measurements predicts an excess of $\approx0.045$~\epp, a $\approx3\sigma$ tension. Either the comet's activity
varied or the marginal Q11 signal is spurious; we therefore regard the Q11 measurement as unconfirmed. The marginal
C/2012~L2
signal comes from a few bins in the first half of the quarter and from times of high solar-wind activity (it falls
to $1.6\sigma$ using only quiet times), L2 is not detected in any other quarter, and its implied \afr\ of
$\sim320$~cm at 12~AU would be extreme; we regard it, like C/2010~G3, as unconfirmed. All other comets give $3\sigma$ limits of
$K_{p,5'}>16.0$--18.5, or \afr\ limits of 1.5--3~cm at 3--4~AU and tens to hundreds of centimeters at 6--13~AU.
Four of the five comets added by extending the selection to $q<15$~AU are not detected: C/2011~Q1 and C/2009~S3 at
7~AU ($K_{p,5'}>17.2$ and $>17.4$--18.1), C/2007~H2 at 8.7~AU, on the detectors for only two days ($>16.0$), and
C/2014~W3 at 11--13~AU ($>17.3$--17.9).

\begin{figure}[p]
\centering
\begin{interactive}{animation}{c_fig5_anim.mp4}
\includegraphics{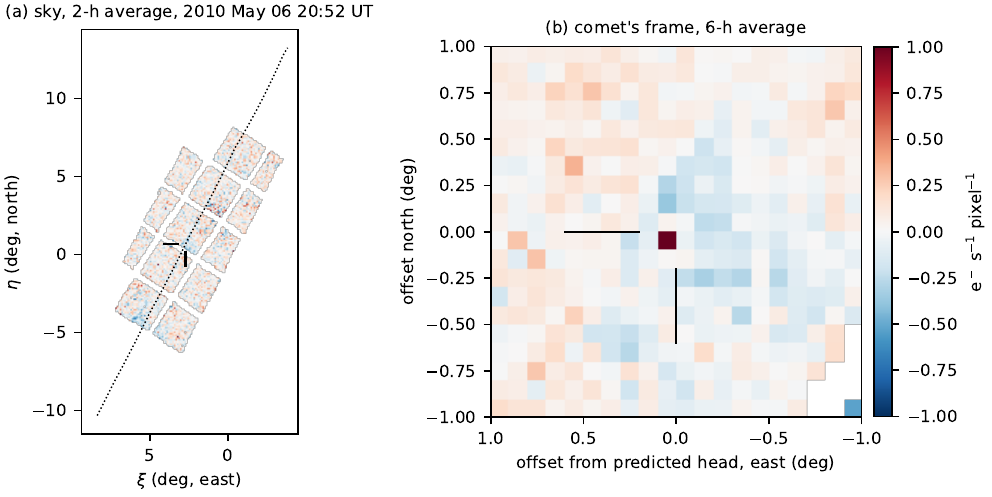}
\end{interactive}
\caption{Comet C/2010~J4 (WISE) on 2010 May 6 20:52 UT (BKJD 490.37), midway through its crossing of the \emph{Kepler}
field. (a) The field, 2-h average, $6'$ cells (tangent plane centered on the field, east left, north up); the tick
marks point to the head position predicted by JPL Horizons, and the dotted line is its track. At $5.8$\dpd\ the comet
moves $\approx0.5^\circ$ during each average and is hard to see. (b) A $2^\circ\times2^\circ$ view in the comet's
frame: the background groups averaged over 6~h by their offsets from the predicted head at each cadence (minus the
median at $30'$--$60'$), which shows the compact coma at the predicted position (red brighter). An animation of this
figure is available. It covers BKJD 488.7--492.2 (2010 May 5 04:48 to May 8 15:48 UT) at 1-h steps and runs for
14~s.}
\label{fig:J4}
\end{figure}

\paragraph{C/2010 J4 (WISE).} J4 passed $0.35$~AU from \emph{Kepler} and crossed the array from south to north in
about two days (BKJD 489.0--491.7, 2010 May 5.5--8.2 UT) at $5.8$\dpd, four to seven days before its discovery in
WISE images on May 12 (Figure~\ref{fig:J4} and its animation).
The excess is consistent from bin to bin (Figure~\ref{fig:lc}), its stack shows a compact head
(Figure~\ref{fig:census}), and its six control tracks lie within $\pm1.4\sigma$. Its low \afr\ of 0.35~cm marks a
weakly active comet that appears bright mainly because it was close: at its distance $5'$ spans 76\,000~km. Its
$K_{p,5'}=14.8$ is 2.9~mag brighter than its JPL total magnitude, as large a difference as for any comet in our
sample (Section~\ref{sec:obs}). The difference is expected from aperture size. The magnitudes reported to the Minor
Planet Center with the astrometry of 2010 May 13--23, after discovery, are 17.5--19.4 (MPC~70177 and 71049); such
magnitudes are measured within a few arcseconds of the nucleus. For a $1/\rho$ coma the flux grows in proportion to
the aperture radius, so a coma with $K_{p,5'}=14.8$ gives $\approx18.5$--19.3 within $5''$--$10''$, consistent with the
reported values and with JPL's magnitude parameters, which are fitted to them. J4 was excluded by an earlier, narrower selection that dropped comets with faint
magnitude parameters ($M_1>18$), which illustrates why we select by orbit.

\begin{figure}[p]
\centering
\begin{interactive}{animation}{c_fig6_anim.mp4}
\includegraphics{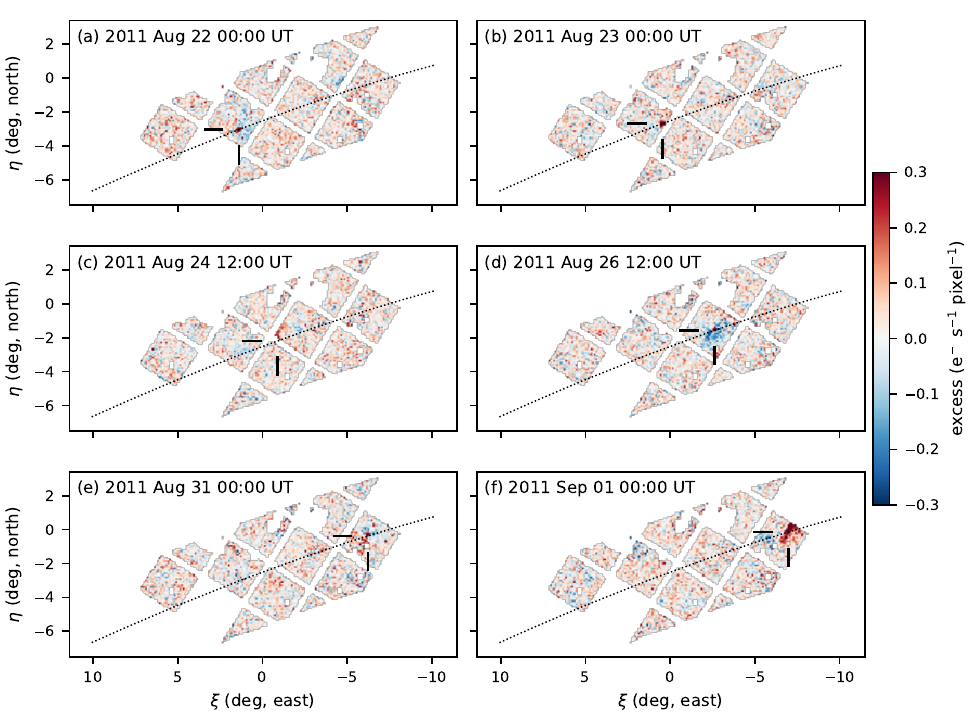}
\end{interactive}
\caption{Comet C/2011~L3 crossing the \emph{Kepler} field in 2011 August (6-h averages, $6'$ cells, red brighter;
tangent plane centered on the field, east left, north up). The two tick marks point to the head position predicted by
JPL Horizons, and the dotted line is its track. Cells with persistent residuals from variable or bright stars are
masked (white). The bluish surroundings of the coma at some times are light from the coma absorbed by the
per-cadence background surface. An animation of this figure is available. It covers BKJD 956.5--974.5 (2011 August
16 00:00 to September 2 23:58 UT) at 2-h steps and runs for 18~s. Left: the part of the field around the comet's
track (6-h averages, $6'$ cells); right: a $4^\circ\times4^\circ$ view centered on the predicted head.}
\label{fig:L3snap}
\end{figure}

\begin{figure}[p]
\centering
\includegraphics{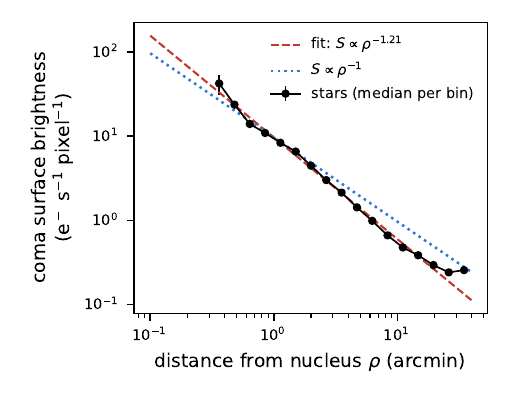}
\caption{Coma surface-brightness profile of C/2011~L3 from the light added to 12\,772 faint stars in Q10, as a
function of each star's distance from the predicted nucleus. Red: power-law fit $S\propto\rho^{-1.21}$; blue:
$\rho^{-1}$.}
\label{fig:L3}
\end{figure}

\paragraph{C/2011 L3 (McNaught).} L3 crossed the southern half of the field in Q10 (BKJD 958--973) at
$0.75$--$1.2$\dpd\ and is, with Y1, the brightest comet recorded ($K_{p,5'}=13.6$). Figure~\ref{fig:L3snap} shows it moving across the
field, and its animation follows it at 2-h steps. It was too close to a CCD gap for the three Q10 full-frame images to be
useful, and its coma is far too thin to dim background stars measurably (optical depth
$\tau\approx$~\afr$/(A\rho)\ll10^{-4}$). Instead, every star near the comet's path acts as a sky probe: stars that the
coma passes over are brightened by the coma's surface brightness at their distance from the nucleus. Stacking
12\,772 stars in Q10 by their distance from the predicted nucleus, each relative to its own light curve 3~days before
and after, gives a coma profile from $0.4'$ to $35'$ (Figure~\ref{fig:L3}) that follows $S\propto\rho^{-1.21}$,
slightly steeper than the $\rho^{-1}$ of steady isotropic outflow, with $S(1')=9.7$~\epp. Part of this steepening
may be instrumental: the pipeline's per-output background fit absorbs some of the coma light, proportionally more at
large $\rho$ where the coma is faint, so the true profile may be closer to $\rho^{-1}$. The profile corresponds to
\afr\ $\approx20$--27~cm at $r=1.93$~AU. Head photometry from the background pixels gives 12~cm. The difference
arises mainly because the head photometry assumes a $1/\rho$ profile (the star-based $S(\rho)$ predicts a head excess
of 1.8~\epp, compared with the 1.1~\epp\ observed); given this and the uncertain slope, we regard \afr\ for L3 as
uncertain by a factor of $\sim2$. The brightness within $5'$, $K_p\approx13.6$, is consistent with visual estimates of
magnitude 13--14 in July--September 2011 \citep{Shanklin2011}.

\paragraph{Distant comets and their dust tails.}
\label{sec:distant}
C/2007~Q3 (Siding Spring) moved at only $\approx0.1$\dpd\ across the western modules in Q9 and Q10, and C/2006~Q1
(McNaught), on its way out from perihelion, at 0.08--0.11\dpd, staying on the detectors for most of each quarter.
Both are active at heliocentric distances where activity is usually driven by species more volatile than water
\citep[e.g.,][]{Jewitt2009,Jewitt2017}. Their stacks (Figure~\ref{fig:census}) show that their light is not a round
coma but a narrow streak extending $\sim30'$ from the head. In the four measurements at $\sigma_{\rm cal}\ge5$ the
streak points along position angle $319^\circ$ and $298^\circ$ (C/2007~Q3, Q9 and Q10) and $273^\circ$ and $265^\circ$
(C/2006~Q1, Q5 and Q6), within $4^\circ$--$16^\circ$ of the projected negative velocity vector (PsAMV; $319^\circ$,
$314^\circ$, $269^\circ$, $269^\circ$) and $42^\circ$--$104^\circ$ from the anti-solar direction. The weaker Q11
measurement of C/2006~Q1, at 10.4~AU, shows no clear direction. For these slow comets the direction of apparent
motion, averaged over each interval (position angles $329^\circ$, $266^\circ$, $333^\circ$ and $277^\circ$), lies
only $8^\circ$--$64^\circ$ from PsAMV, so the direction alone does not distinguish a dust tail from a smear along the
track. Two properties argue against an artifact. The streaks are one-sided: within $\pm30^\circ$ of the streak
direction, the stacked excess $10'$--$36'$ from the head has signal-to-noise ratio 2.6--8.3, compared with $-1.8$ to
$+0.2$ in the opposite direction. They also lie on the side toward which the comet was moving across the sky (a motion dominated by parallax from
\emph{Kepler}'s orbit, not by the comet's own orbital motion), whereas the
running-median baseline can only lower the signal along the track, and the ephemerides of these well-observed comets
are accurate to much better than our $4'$ cells. We therefore interpret the streaks as dust tails of large, slow
grains, which at these distances lie close to the negative velocity direction, but higher-resolution images would
be needed to confirm them. Light from these tails within $8'$ is
included in the head photometry, so their \afr\ values are upper bounds on that of the coma alone.

\section{General comet detection}
\label{sec:generalcomet}

\subsection{Blind search for moving diffuse sources}
\label{sec:blind}

The comet images discussed above were obtained by stacking \emph{Kepler} data along the predicted sky tracks of known comets, which provides a signal-to-noise improvement over fixed time images. To evaluate whether these comets could have been detected without any knowledge of their orbits, we searched every quarter for moving diffuse sources of arbitrary direction.  For this test, we computed 
the temporal baseline of each pixel group from a regression on its output's mean
level, that level squared, and three slowly varying principal components of the output, which removes instrumental
drifts while keeping multi-day signals. The resulting background layer was binned into $20'$ cells
and 6-h bins, normalized by each cell's robust noise and by each frame's robust spread (which suppresses field-wide
disturbances), and high-passed in time by removing each cell's running median over $\pm3$~days. Frames were then
shifted along trial constant velocities and summed (``shift and stack'', or synthetic tracking): $\approx1000$
velocities up to 3\dpd\ in 2-day windows, and $\approx250$ velocities up to 0.5\dpd\ in 6-day windows. Peaks with
stacked significance $S\ge6$ that exceed the best stationary stack by $\ge2$ and do not recur at the same place in
separate epochs were merged into candidates, checked against the ephemerides of all known comets, and inspected in
fixed-sky ``vetting strips''. A null test with the time order of frames shuffled gives no peak above $S=6.6$.

The search recovers C/2011~L3 ($S\approx12$ in several independent windows, within $0.05^\circ$--$0.18^\circ$ of
the predicted position, at the correct speed), C/2009~Y1 ($S=12$), C/2007~Q3 in Q9 ($S=9.2$, $0.12$\dpd), and
marginally C/2010~G1 ($S=6.2$). It also detects the K5 ion tail as a set of fast-moving structures in Q5. It misses
the fainter comets detected by targeted photometry, as expected given the trials penalty of a blind search, and also
the bright but fast C/2010~J4 ($5.8$\dpd), which lies outside its velocity range. A separate search of Q5 with 3-h
frames, 1-day windows and speeds up to 8\dpd\ does not recover J4 either: in the $20'$ cells of the search its compact
coma is diluted and smeared over $\approx45'$ per frame, so that each frame carries only $\approx1.7\sigma$, and even a
stack that follows the exact ephemeris reaches only $S=5.9$ in a 1-day window and 7.5 over the whole 2.4-day crossing,
at the threshold before any trials penalty. Comets that move faster than $\approx3$\dpd, which must pass within
$\sim0.5$~AU of \emph{Kepler}, are therefore beyond the reach of the blind search, even when bright. All
strong unidentified candidates were traced to artifacts: a field-wide brightening on 2012 November 13 coincident
with the arrival of an interplanetary CME, thermal settling after spacecraft rolls, edges of CCD gaps, and stationary pairs of
positive and negative residuals of bright or variable stars. One positive, stationary diffuse transient (right
ascension $295.3^\circ$, declination $42.3^\circ$, BKJD 308.5--309.3) remains unexplained but does not move. No new
comets were found.

Injection tests (synthetic $1/\rho$ comae on straight tracks; Table~\ref{tab:inject}) give a completeness of about
50\% at $K_{p,5'}\approx15.5$--16 for $0.3$--$1$\dpd, lower at 2.5\dpd, where a comet crosses a $20'$ cell in less
than a bin, and none at 0.1\dpd, where the $\pm3$-day high-pass removes slow movers. C/2007~Q3, recovered at 0.12\dpd\ although
its $K_{p,5'}=16.3$ is fainter than the 50\% level, is probably an exception because its light extends $\sim30'$
along a dust tail (Section~\ref{sec:results}) rather than forming a compact coma like the injected ones. With two injections per entry
these numbers are indicative only; slowly moving comets would require a search with a longer high-pass and longer
windows.

\begin{table}[t]
\centering
\caption{Blind-search completeness: recovered/injected (two injections per entry).}
\label{tab:inject}
\begin{tabular}{ccrrrr}
\toprule
$S_1$ (\epp) & $K_{p,5'}$ & 0.1\dpd & 0.3\dpd & 1\dpd & 2.5\dpd\\
\midrule
0.25 & 17.0 & 0/2 & 0/2 & 0/2 & 0/2\\
0.5 & 16.2 & 0/2 & 1/2 & 1/2 & 0/2\\
1.0 & 15.5 & 0/2 & 2/2 & 1/2 & 1/2\\
2.0 & 14.7 & 0/2 & 2/2 & 2/2 & 1/2\\
\bottomrule
\end{tabular}
\end{table}

\subsection{Observability of comets with \emph{Kepler}}
\label{sec:obs}
\begin{figure}[p]
\centering
\includegraphics{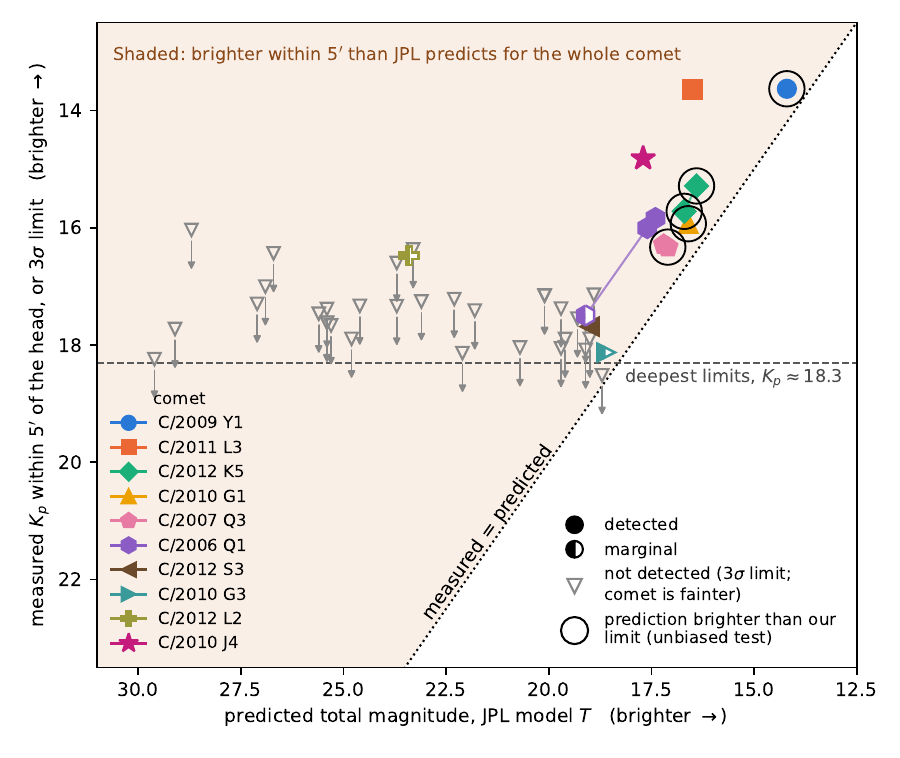}
\caption{Measured magnitude within $5'$ of the head versus the JPL model total magnitude $T$; brighter is up and to the
right on both axes. Each comet has its own symbol and color; filled: detections; half-filled: marginal
measurements; lines join measurements of the same comet. Gray triangles with arrows are $3\sigma$ limits for undetected comets, which are fainter than the
triangle. The dotted line marks $K_{p,5'}=T$: light within $5'$ cannot exceed a comet's total light, so if the JPL
magnitudes were correct, detections would lie on or below it (unshaded region). Rings mark the five measurements for
which $T$ is brighter than our detection limit, which therefore test the predictions without selection bias. The
dashed line marks the deepest $3\sigma$ limits, $K_{p,5'}\approx18.3$.}
\label{fig:obs}
\end{figure}

\begin{figure}[p]
\centering
\includegraphics{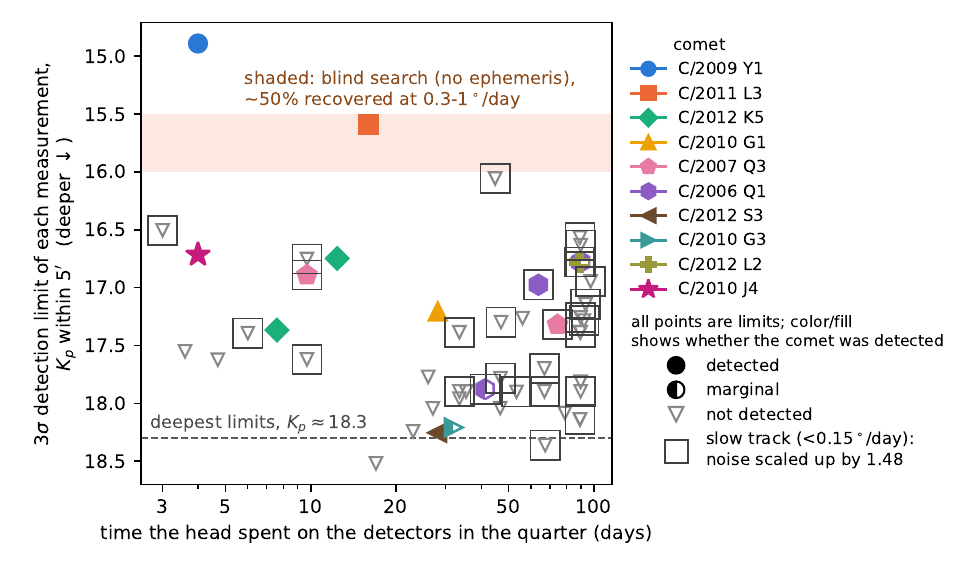}
\caption{Depth of the targeted photometry: the $3\sigma$ limit of every measurement, detected (colored) or not
(gray), versus the time the comet's head spent on the detectors in that quarter; comet symbols as in
Figure~\ref{fig:obs}. Open squares mark slow tracks
($<0.15$\dpd), whose noise is scaled up by 1.48 to allow for correlated noise. The shaded band shows where the blind
search, without ephemerides, recovered about half of the injected synthetic comets moving at 0.3--1\dpd.}
\label{fig:depth}
\end{figure}

Three factors set what \emph{Kepler} can detect (Figures~\ref{fig:obs} and \ref{fig:depth}).

\emph{Sampling and surface brightness.} Background groups sample only $\approx0.4\%$ of the sky on a $2'$ grid, so
the head of a comet rarely falls on a sampled pixel and \emph{Kepler} measures the coma's extended envelope, from a
few to tens of arcminutes, rather than its central condensation. The relevant quantity is therefore $S_1$, or
equivalently the magnitude within a few arcminutes. The total magnitudes predicted by JPL's simple magnitude laws are
a poor guide to this quantity (Figure~\ref{fig:obs}). Light within $5'$ cannot exceed a comet's total light,
so if the JPL magnitudes $T$ were correct the detections would lie on or below the line $K_{p,5'}=T$; instead all lie
0.6--2.9 magnitudes above it. Because $K_{p,5'}$ is a lower bound on a comet's total light, the detected comets
are at least this much brighter than predicted. This comparison needs no model of the outer coma, which matters
because a $1/\rho$ coma has no finite total magnitude without an outer cutoff. $K_{p,5'}$ itself depends on the
assumed profile, but only weakly: for $S\propto\rho^{-\alpha}$ with $\alpha=0.7$--1.5 it changes by at most 0.11~mag,
and in the sense that makes the comet brighter than we infer. The per-output background fit, which absorbs some of the
coma light (C/2011~L3, Section~\ref{sec:results}), also makes our magnitudes too faint. Much of the excess reflects selection: for nine of the fourteen detected or marginal measurements
$T$ is fainter than our $3\sigma$ limit, so those comets could only be detected by being brighter than predicted. The
five measurements for which $T$ is brighter than our limit, and which are therefore free of this bias (C/2009~Y1,
C/2010~G1, C/2007~Q3 in Q9 and C/2012~K5 in two quarters), were all detected and are at least 0.6--1.1 magnitudes
brighter than $T$. A few tenths
of a magnitude of this may come from the difference between the \emph{Kepler} band and visual magnitudes; the rest
is consistent with JPL's magnitude laws being fitted to small-aperture astrometric magnitudes. No undetected comet
had a JPL magnitude brighter than its limit, so the non-detections do not test the predictions.

\emph{Time on the detectors and angular speed.} The $3\sigma$ limits range from $K_{p,5'}=14.9$ to 18.5 (80\% between
16.6 and 18.1; Figure~\ref{fig:depth}), set by the noise of each measurement rather than by a single sky-brightness
threshold. The noise falls roughly as the square root of the number of independent 0.5-day bins, so a comet crossing
in four days, like C/2009~Y1, reaches only $K_{p,5'}\approx14.9$, while comets on the detectors for 20--40 days reach
$\approx18$. Longer still does not help: the comets that stay for 60--90 days are the slow, distant ones, and
separating them from static sky structure requires baselines of up to ten days, over which variable stars and slow
instrumental drifts add correlated noise (the factor 1.48 of Section~\ref{sec:census}), so their limits are 16.4--18.1.
$K_{p,5'}=18$ corresponds to a $1/\rho$ coma with $S=0.1$~\epp\ at $1'$ and $\approx0.02$~\epp, or $\approx10^{-4}$ of the
sky, at $5'$.

\emph{Blind versus targeted.} Without an ephemeris, the trials penalty of a velocity search raises the threshold by
$\approx1.5$--2.5~mag, and the blind search covers only speeds up to 3\dpd. A comet bright enough to be found blindly, $K_{p,5'}\lesssim15.5$, would have been well within
reach of the ground-based surveys of 2009--2013 at declination $+45^\circ$, so finding no new comets is the expected
outcome. \emph{Kepler}'s value for comets lies instead in high-cadence monitoring of known objects --- for example
outbursts, and solar-wind interaction as in C/2009~K5 --- and in diffuse structures, such as ion and dust tails, that
are hard to measure from the ground.

\section{Discussion}

\paragraph{Brightness of C/2009 K5.} Predictions differed widely from the comet's observed brightness. JPL's
total-magnitude parameters ($M_1=11.9$, $k_1=5$; fitted to astrometric CCD magnitudes) give $T\approx13.3$ for
mid-April 2010, and the Minor Planet Center (MPC) ephemeris gave $\approx9.6$. Visual observers reported the comet at
around magnitude 8 near its perihelion (1.4~AU) in late April 2010, fading by mid-May
(\citealt{Shanklin2010}; British Astronomical Association comet section, 84 observations). The comet was therefore
substantially more active than predicted, consistent with a well-developed plasma tail. Because the nucleus never
fell on the detectors, \emph{Kepler} provides no independent coma photometry of K5.

\paragraph{Sensitivity of \emph{Kepler} background photometry.} The noise in $10'\times10'$ cells and 3-h bins is
$\approx0.07$--0.08~\epp, i.e. $\approx3\times10^{-4}$ of the sky. \emph{Kepler}'s sky (22--22.7~mag\,arcsec$^{-2}$ in
the \emph{Kepler} band) has no airglow and is about a magnitude or more darker than a dark ground-based site in a
similarly broad red band, which lowers the photon noise by a factor of $\approx1.5$--2. The measured noise, however,
is about five times the photon noise of the $\sim100$ pixels sampled in such a cell ($\approx0.014$~\epp), so the
limit is set by systematics rather than by photons. The decisive advantages are therefore stability and continuity:
there are no airglow fluctuations, changes of atmospheric extinction, weather or daylight, and the same pixels are
recorded every 29.4~min for years, so static sky structure and flat-field errors cancel when each pixel is compared
with its own history.

\begin{figure}[p]
\centering
\includegraphics{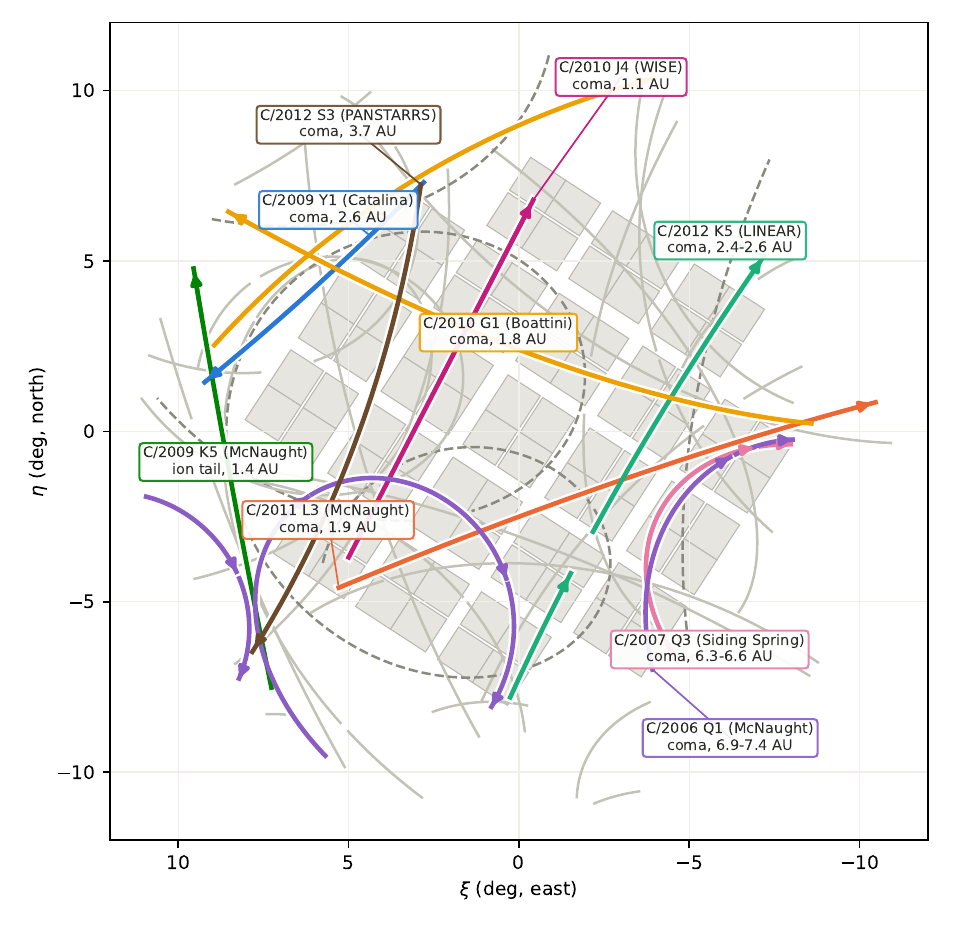}
\caption{Paths of known comets across the \emph{Kepler} field while \emph{Kepler} observed (positions as seen from
\emph{Kepler}, JPL Horizons), over the CCD layout. Colored and labeled (type of detection, heliocentric distance): comets
detected in background pixels; dashed: marginal
(C/2010~G3, C/2012~L2); gray: not detected. C/2006~Q1, marginal in Q11, is drawn as detected. Paths are shown for the whole time each comet was near the field, including quarters in which it was not
detected. The curvature of distant comets' paths is parallax from \emph{Kepler}'s orbit.}
\label{fig:paths}
\end{figure}

\paragraph{Effects on \emph{Kepler} light curves.} The standard processing pipeline subtracts from each target aperture a background
estimated by a low-order polynomial fit over each CCD output \citep{Jenkins2010,Twicken2010}. This removes smooth,
field-wide diffuse light: in the companion paper we show that Thomson-scattered solar wind, at 0.02--0.1~\epp,
changes PDCSAP light curves of $K_p=15$--16 stars by $\lesssim1$~ppm. Comets differ because their light is
structured on scales of arcminutes, which the background fit cannot follow. The K5 ion tail added a uniform pedestal
of 1--4~\epp\ to the apertures it crossed; for faint stars ($K_p\approx15$--16) that is up to $\approx0.4\%$ for
$\approx1.5$~h, which is how the tail was found (Section~2.2). Averaged over all stars on the tail the change is
smaller, $\sim10^{-4}$, partly because the pipeline's per-output background fit takes up some of the tail light;
the same fit then over-subtracts slightly elsewhere on the output, leaving small negative patches in the light curves
of stars near the tail. A coma can do the same: with the C/2011~L3 profile
($S\approx4.5$~\epp\ at $2'$ and $\approx0.5$~\epp\ at $10'$), a 12-pixel aperture of a $K_p=15$ star gains
$\approx0.5\%$ when the nucleus passes within $2'$ and $\approx0.05\%$ within $10'$, for one to several hours. Such
events are rare, brief and confined to the stars along a comet's path. Apparent brightenings of faint stars that
coincide with the comet crossings listed in Table~\ref{tab:census} and Figure~\ref{fig:paths}, or with the K5 tail
in 2010 April, should be treated with caution. Otherwise, diffuse foreground light does not affect \emph{Kepler}
photometry at a level that matters for transit or variability studies.

\paragraph{Related phenomena.} The same searches revealed $\sim13$ brief streak events over the mission, each
3--$10^\circ$ long, 4--$10'$ wide and usually confined to a single 29.4-min cadence. One possible explanation is
sunlight scattered by small particles near the spacecraft, which would appear strongly out of focus; we have not
investigated them further.

\paragraph{Outlook.} The same methods could be applied to the K2 mission (2014--2018), whose fields along the ecliptic were
crossed by many more comets, and to the full-frame images of the Transiting Exoplanet Survey Satellite (TESS), which
sample every pixel and have already been used for cometary science \citep[e.g.,][]{Farnham2019,Farnham2021}.

\section{Conclusions}
\begin{enumerate}
\item \emph{Kepler}'s background pixels are a sensitive, under-used record of faint diffuse light, suitable for
systematic searches for moving sky features, including comets.
\item A moving band in Q5 \emph{Kepler} photometry is the plasma tail of comet C/2009~K5 (McNaught), traced from
$0.9^\circ$ to $\approx13^\circ$ from the nucleus at surface brightnesses of 26.7--30~mag\,arcsec$^{-2}$.
\item The tail axis is offset $4.4^\circ$--$5^\circ$ from the anti-solar direction, implying a solar-wind speed of
$\approx500$--750~km\,s$^{-1}$ at the comet, and the tail widens from $6'$ to $\approx20'$.
\item A disconnection event on 2010 April 17--19 is resolved at 30-min cadence; the detached tail receded at
$\approx400$~km\,s$^{-1}$. The timing and direction of a CME launched on 2010 April 13 are consistent with a flank
encounter shortly before the disconnection, although a current-sheet or stream-interface crossing cannot be excluded.
\item Uniform photometry of the 21 known comets whose heads fell on the \emph{Kepler} detectors, calibrated with
displaced control tracks, detects eight comae, including C/2010~J4 a few days before its discovery, and C/2007~Q3 and
C/2006~Q1 at 6--7.5~AU, whose light extends in one-sided streaks that we interpret as dust tails. The detected comets are at least 0.6--2.9 magnitudes brighter than JPL predicts,
partly because of selection (Section~\ref{sec:obs}). The coma profile of C/2011~L3 is measured from the light it added to
faint stars.
\item Targeted photometry reaches $K_p\approx18$ within $5'$ for comets that remain on the detectors for three to six
weeks; a blind shift-and-stack search is roughly 50\% complete at $K_{p,5'}\approx15.5$--16 for speeds of
0.3--1\dpd, cannot follow fast comets such as C/2010~J4, and finds no new comets.
\end{enumerate}

\section*{Data and code availability}
The \emph{Kepler} background-pixel files and light curves are available from MAST, comet ephemerides from JPL
Horizons and the Small-Body Database, and CME catalogs from the CDAW Data Center and HELCATS. The reduced
background archive and the code used for the analyses, figures and animations are available on Zenodo:
\dataset[doi:10.5281/zenodo.24182076]{https://doi.org/10.5281/zenodo.24182076}. All the \emph{Kepler} long-cadence data used in this paper can be found in MAST:
\dataset[doi:10.17909/T9488N]{https://doi.org/10.17909/T9488N}.

\begin{acknowledgments}
This paper includes data collected by the \emph{Kepler} mission and obtained from the MAST data archive at the Space
Telescope Science Institute (STScI). Funding for the \emph{Kepler} mission is provided by the NASA Science Mission
Directorate. STScI is operated by the Association of Universities for Research in Astronomy, Inc., under NASA contract
NAS~5--26555. We used the JPL Horizons system and Small-Body Database. The SOHO/LASCO CME catalog is generated and
maintained at the CDAW (Coordinated Data Analysis Workshop) Data Center by NASA and The Catholic University of America
in cooperation with the Naval Research Laboratory; SOHO is a project of international cooperation between ESA and
NASA. The HELCATS project received funding from the European Union's Seventh Framework Programme (FP7/2007--2013)
under grant agreement no.~606692. We thank the observers of the British Astronomical Association comet section.
Analysis and narration assisted by Claude (Anthropic).
\end{acknowledgments}

\facilities{Kepler}

\software{NumPy \citep{Harris2020}, Matplotlib \citep{Hunter2007}}

\clearpage

\begin{figure}[p]
\centering
\includegraphics{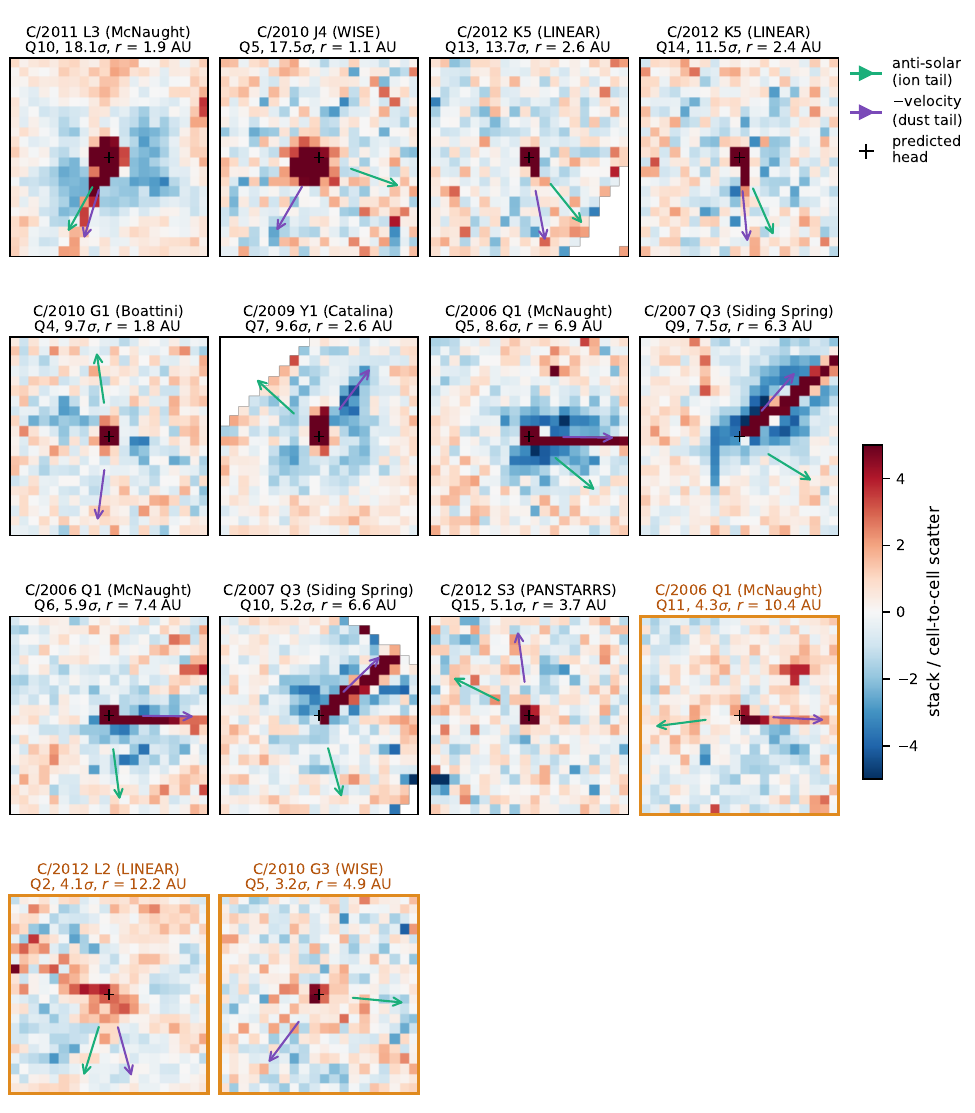}
\caption{Co-moving stacks of every detected and marginal measurement (marginal ones framed in orange), in order of
calibrated significance $\sigma_{\rm cal}$; titles give the quarter, $\sigma_{\rm cal}$ and heliocentric distance. Each
panel shows $80'\times80'$ around the predicted head (cross; east left, north up, $4'$ cells), in units of the stack's
cell-to-cell scatter. Arrows show the anti-solar direction (expected ion tail, green) and the projected negative
velocity direction (expected dust tail, purple). Blue halos are side-lobe artifacts (see text). The head light curves are in Figure~\ref{fig:lc}.}
\label{fig:census}
\end{figure}

\clearpage
\appendix
\renewcommand{\thefigure}{A\arabic{figure}}\setcounter{figure}{0}

\section{Head light curves}\label{app:lc}
\begin{figure}[p]
\centering
\includegraphics{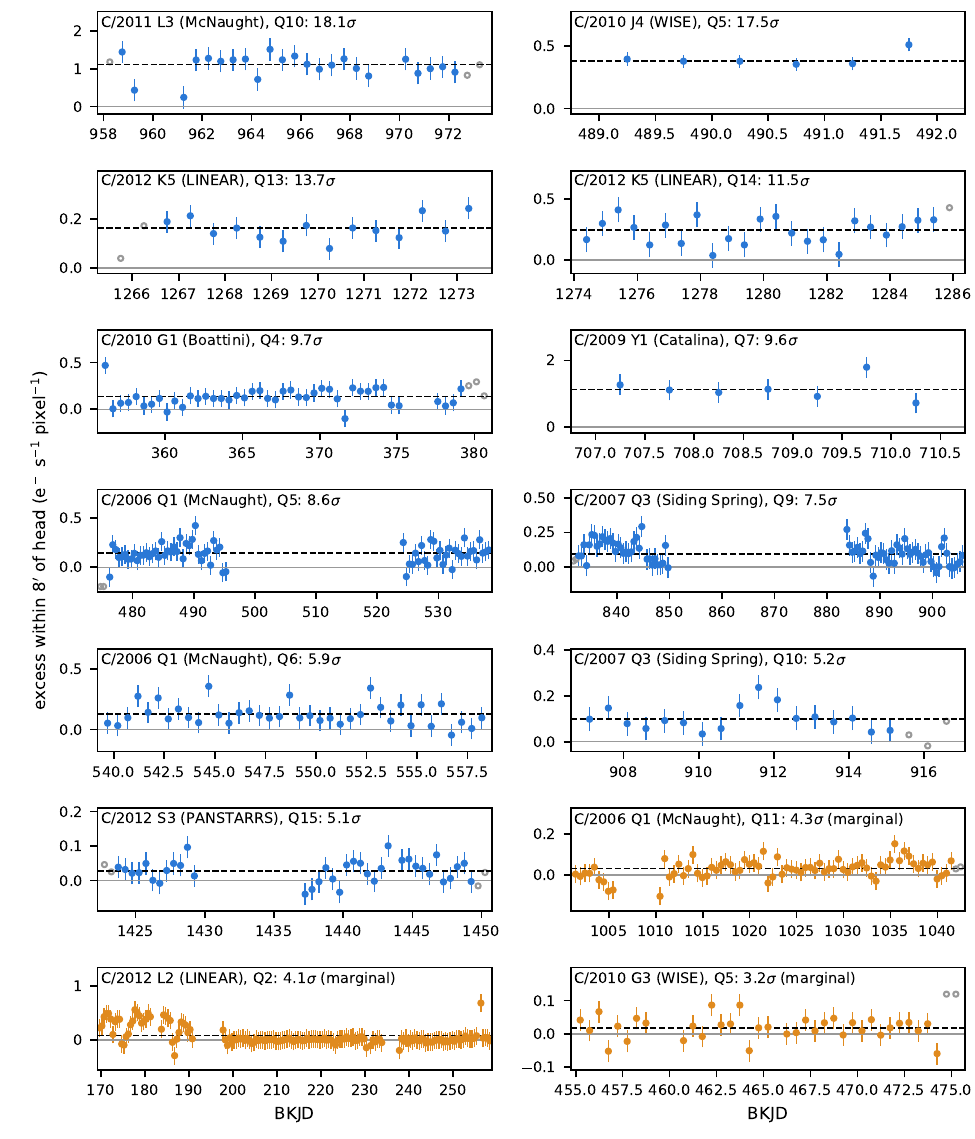}
\caption{Head light curves of the measurements in Figure~\ref{fig:census} (detections blue, marginal orange), in the
same order. Each point is a 0.5-day average of the excess within $8'$ of the predicted head minus a $16'$--$32'$ ring,
with error bars from the scatter between bins; the dashed line is the weighted mean and the gray line zero. Open gray
points are bins before the head reached or after it left the detectors, which are not used.}
\label{fig:lc}
\end{figure}

Figure~\ref{fig:lc} shows the head light curves of the detected and marginal measurements listed in
Table~\ref{tab:census} and shown in Figure~\ref{fig:census}.

\end{document}